\documentclass[]{spieman}  

\usepackage{amsmath,amsfonts,amssymb}
\usepackage{graphicx}
\usepackage{xcolor}

\usepackage{framed}
\usepackage{lipsum}

\usepackage{adjustbox}
\usepackage{soul}

\usepackage{subcaption}
\usepackage[colorlinks=true, allcolors=blue]{hyperref}

\def\xray{X-ray}

\title{In orbit operation of Resolve Filter Wheel and MXS}

\author[a]{Russell F. Shipman}
\author[b]{Shunji Kitamoto}
\author[a,c]{Rob Wolfs}
\author[a]{Elisa Costantini}
\author[d]{Megan E. Eckart}
\author[e]{Carlo Ferrigno}
\author[f]{Ludovic Genolet}
\author[a]{Nathalie Gorter}
\author[a]{Martin Grim}
\author[a]{Jan Willem den Herder}
\author[g]{Caroline A. Kilbourne}
\author[g]{Maurice A. Leutenegger}
\author[a]{Erik van der Meer}
\author[h]{Misaki Mizumoto}
\author[g]{F. Scott Porter}
\author[e]{St\'ephane Paltani}
\author[b]{Makoto Sawada}
\author[i]{Simon Strotmann}
\author[j]{Masahiro Tsujimoto}
\author[a]{Cor P. de Vries}

\affil[a]{SRON: Space Research Organisation Netherlands, Niels Bohrweg 4 2333 CA Leiden, Netherlands}
\affil[b]{Department of Physics, Rikkyo University, 3-34-1 Nishi Ikebukuro, Toshima-ku, Tokyo 171-8501, Japan}
\affil[c]{DIFFER: Dutch Institute for Fundamental Energy Research,  De Zaale 20, 5612 AJ Eindhoven, Netherlands}
\affil[d]{Lawrence Livermore National Laboratory, 7000 East Avenue, Livermore CA 94550, USA}
\affil[e]{University of Geneva, Department of Astronomy, Ch. d’Ecogia 16, 1290, Versoix, Switzerland}
\affil[f]{Observatoire de Gen\`eve Switzerland, Chemin Pegasi, 51
1290 Versoix, Switzerland}
\affil[g]{NASA Goddard Space Flight Center, 8800 Greenbelt Rd., Greenbelt, MD 20771  USA}
\affil[h]{University of Teacher Education Fukuoka, 1-1 Akama-bunkyo-machi, Munakata, Fukuoka 811-4192, Japan}
\affil[i]{European Space Research and Technology Ctr. Keplerlaan 1, 2201 AZ Noordwijk Netherlands}
\affil[j]{Japan Aerospace Exploration Agency (JAXA), Institute of Space and Astronautical Science (ISAS), 3-1-1 Yoshino-dai, Chuo-ku, Sagamihara, Kanagawa 252-5210, Japan}

\begin{document}

\maketitle

\begin{abstract}
The Resolve soft \xray{} spectrometer is the high spectral resolution microcalorimeter spectrometer for the XRISM mission.  In the beam of Resolve there is a filter wheel containing \xray{} filters.  In the beam also is an active calibration source (the modulated \xray{} source (MXS) that can provide pulsed \xray s to facilitate gain calibration. 

The filter wheel consists of six filter positions.  Two open positions, one $^{55}$Fe source to 
aid in spectrometer characterization during the commissioning phase, and three transmission filters: a neutral density filter, an optical blocking filter, and a beryllium filter.

The \xray{} intensity, pulse period, and pulse separation of a MXS are highly configurable.  
Furthermore, the switch--on time is synchronized with the spacecraft's
internal clock to give accurate start and end times of the pulses.  

One of the issues raised during ground testing was the susceptibility of a MXS at high voltage to ambient light.  
Although measures were taken to mitigate the light leak, the efficacy
of those measures must be verified in orbit.  
Along with an overview of issues raised during ground testing,
this article will discuss the calibration source and the filter performance in--flight and compare with
the transmission curves present in the Resolve calibration
database.

\end{abstract}


\section{Introduction}
\label{sec:intro}
The Resolve\cite{Ishisaki2022full} soft \xray{} spectrometer
on the X-ray Imaging and Spectroscopy Mission (XRISM) \cite{tashiro2022}
will perform high-resolution X-ray spectroscopy of astrophysical objects.  
The Resolve microcalorimeter array\cite{kelley_full} is a 6$\times$6 array of square pixels kept at a temperature of 50 mK, mounted inside a Dewar\cite{kelley_full}.   The operating temperature is maintained by an adiabatic demagnetization refrigerator (ADR) cooler.  The gains, or energy response, of the pixels are highly dependent on subtle changes in temperature and will change throughout an orbit and between ADR re-cycles\cite{porter_full}.

In order to achieve the spectral resolution requirement of Resolve, periodic observations of \xray{}s of known energy are required.  This calibration is partially satisfied by having one of the 36 pixels permanently behind a radioactive $^{55}$Fe source: the so-called cal-pixel.  However, the gain of the full array still varies relative to the cal-pixel.  Hence calibration sources that can be placed in the line of sight of the detector are necessary. 
 
In the optical beam of Resolve, there is a filter wheel (FW) to aid in the calibration of the Resolve array.  The Resolve FW is based on the Hitomi FW and calibration sources which are described in [\citenum{10.1117/12.855880}]. To keep development costs low, very little of the design of the filter wheel was changed, and hence the layout of the filters and the calibration sources remains largely the same.  

The filter wheel consists of six filter positions: two open positions, three transmission filters, and one position containing radioactive $^{55}$Fe.  The $^{}55$Fe filter emits Mn \xray{}s.  Transmission filters will be used to uniformly attenuate all \xray{}s or will be needed to allow brighter sources \xray{} to be observed by attenuating the incident beam. Different options (neutral density, band-pass filters) are available, which can be chosen according to the scientific needs. Independent of filter choices, another calibration source, the modulated X-ray source is placed beneath the filter wheel and the Dewar to illuminate the detector array.

On 6 September (UTC) 2023, XRISM was successfully launched into low Earth orbit, and initial commissioning began shortly after\cite{maeda_full}.  
For ground calibration purposes and protection of the detector during early in orbit operations, Resolve is equipped with a partially transparent 270 $\mu m$ thick Be lid called the gate valve \cite{10.1117/1.JATIS.7.2.028005}. 
The gate valve did not open on 6 November and subsequent attempts.  Together with science, this impacts the detector drift monitoring/correction strategy (see [\citenum{porter_full}] and [\citenum{sawada_full}]) as well as the general calibration of the energy scale\cite{megan_full}. 

This article reviews the performance of the FW as measured during the commissioning phase of operations just after launch.  The FW filters and mechanism are described in Section \ref{sec:FW}, the measured transmission through various filters is shown in Section \ref{sec:transmission}, the $^{55}$ Fe position is discussed in Section \ref{sec:fe55}. In \ref{sec:trend}, we discuss how the FW motors will be monitored during the XRISM mission.    {The performance of the modulated} \xray{}  {sources (MXSs) during} commissioning is discussed in Section \ref{sec:mxs}.  In Section \ref{sec:lightleak} some of the observations of the  {MXSs} during laboratory ground tests are compared with the in orbit performance.  

\begin{figure}[ht]
\centering
  
\includegraphics[width=0.9\textwidth]{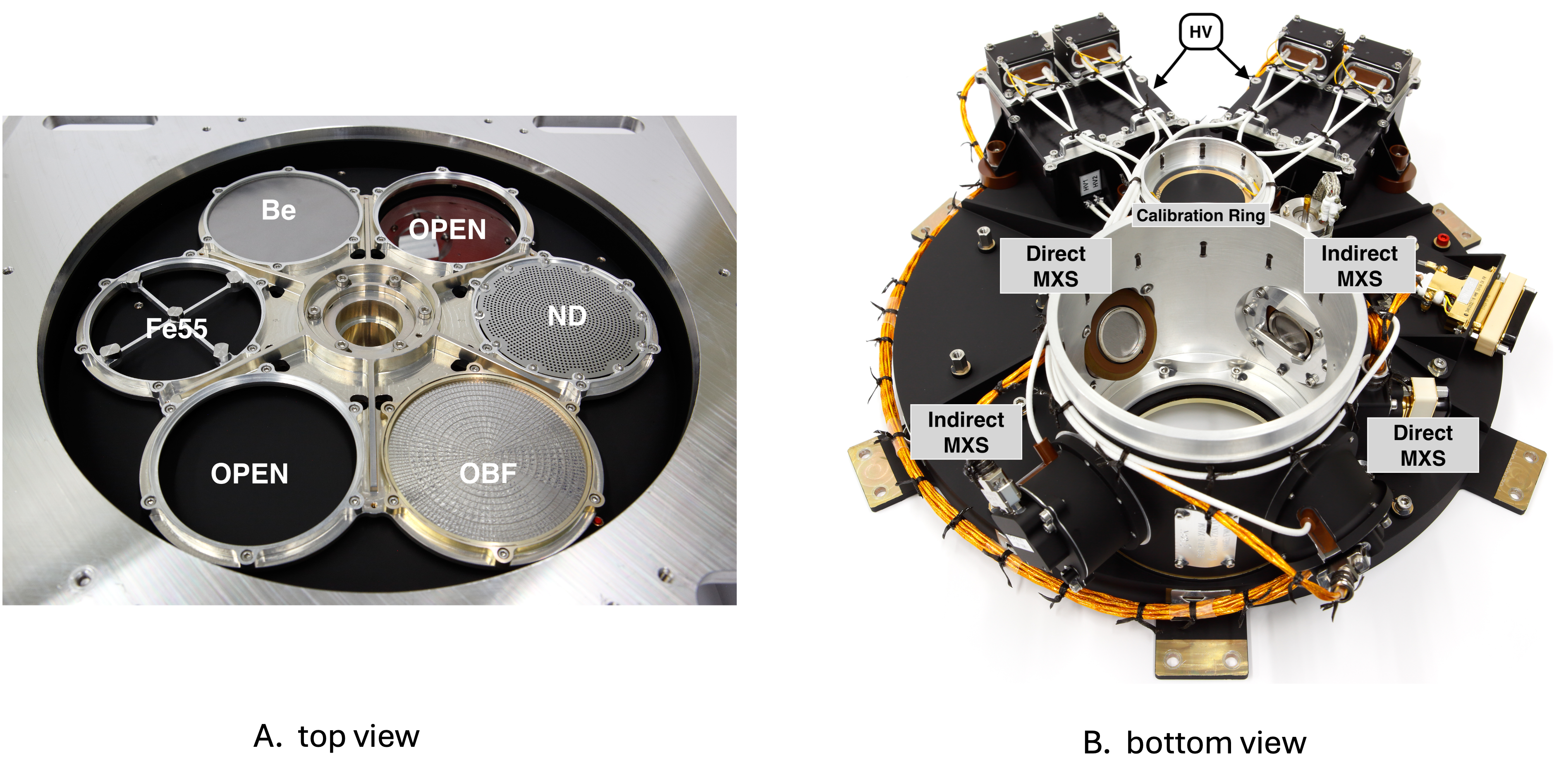}
\caption{On the left (A) is a top view of the FW, the different filters are identified: a beryllium filter (Be), a neutral density filter (ND) and an optical blocking filter (OBF).  Also shown are the open positions and the \xray{} source ($^{55}$Fe).   On the right (B) is a bottom view of the FW.  Here the calibration ring is seen with the layout of the four  {MXSs}.  Also, the prime and redundant high voltage (HV) power supplies are indicated.}
\label{fig:test}
\end{figure}

\section{Filter Wheel} 
\label{sec:FW}

The FW is mounted about 90 cm above the detector array outside of the Dewar.  The wheel can rotate different filters into the optical path of Resolve.  Fig. \ref{fig:test}A shows the six possible positions: two open filters, three transmission filters, and one \xray{} source ($^{55}$Fe).
The filters aid in the observation and calibration of astronomical sources.   To help in observing bright \xray{} sources, the neutral density (ND) filter blocks roughly 75\% of all \xray{}s. An optical blocking filter (OBF)  reduces the UV radiation of optically bright \xray{} sources and a beryllium (Be) filter is present to reduce low energy \xray{}s.

\subsection{transmission filters}
\label{sec:transmission}
\begin{figure}[ht]
\begin{center}
\includegraphics[width=0.70\textheight]{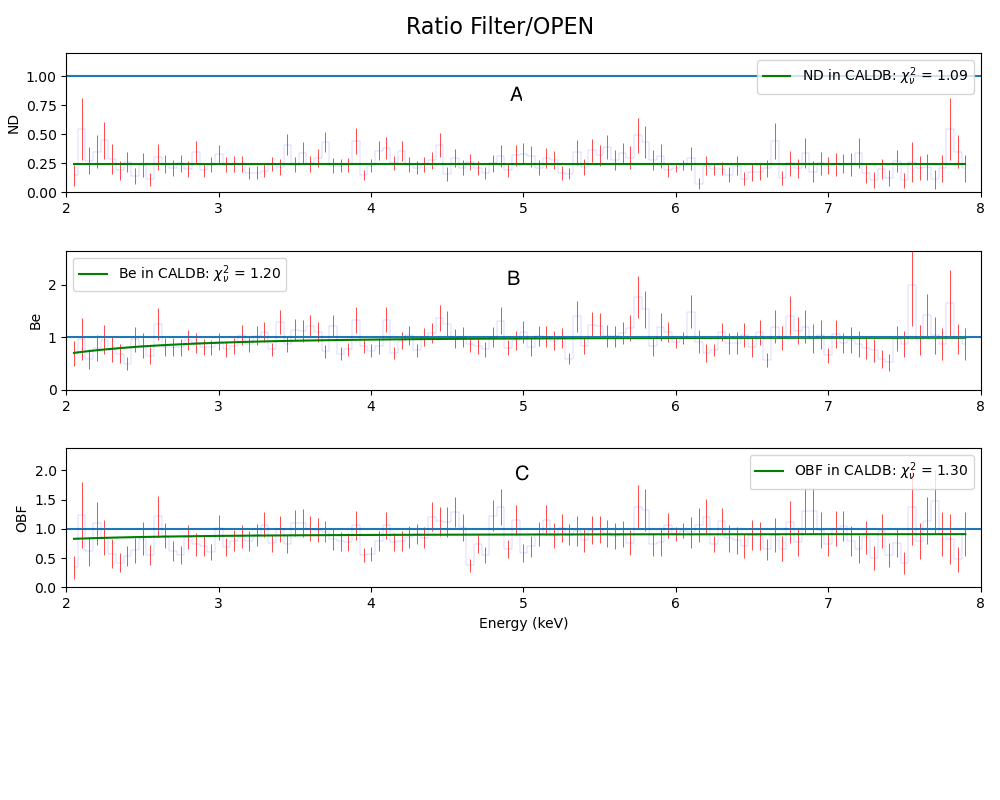}
\end{center}
\vspace{-0mm}
\caption{Transmission of ND  {filter (panel A), Be filter (panel B) and OBF (panel C)} filters in 50 eV bins.  Measured against PKS0745-19. Faded blue lines and red bars are the transmission data and uncertainty for energies between 2 and 8 keV.  The expected transmission is shown in green, the blue line indications 100\% transmission.  Each panel also lists the reduced $\chi^2$ compared with the transmission present in the CALDB for energy ranges between 2 and 10 keV in agreement with ground testing. \label{fig:filtertrans}}
\end{figure}

During commissioning, the transmission of the transmission filters was measured using the source PKS0745-19 (9318 seconds in open position). It was observed in the OBF for 6796 seconds, in the Be filter for 13476 seconds, and in the ND for 11932 seconds. The measured transmissions are shown in red in Fig. \ref{fig:filtertrans}.  The solid green curves in the figures are the transmission trends present in the HEASARC calibration database (CALDB).  The CALDB curves for the OBF and Be filters were created from laboratory measurements at a much higher energy resolution than that presented here. The ND transmission was not measured on-ground, but is based on "as designed" expectations. The blue line indicates 100\% transmission and is present for reference.

In the gate valve closed case, it is not possible to test the transmission below about 2 keV.  This is not an issue for the ND filter, since that filter is designed to be energy independent.  Both the Be and the OBF agree with the expected transmission above 2 keV.

\subsection{\xray{} source position }
\label{sec:fe55}

\begin{figure}[ht]

\begin{center}
\includegraphics[width=0.60\textheight]{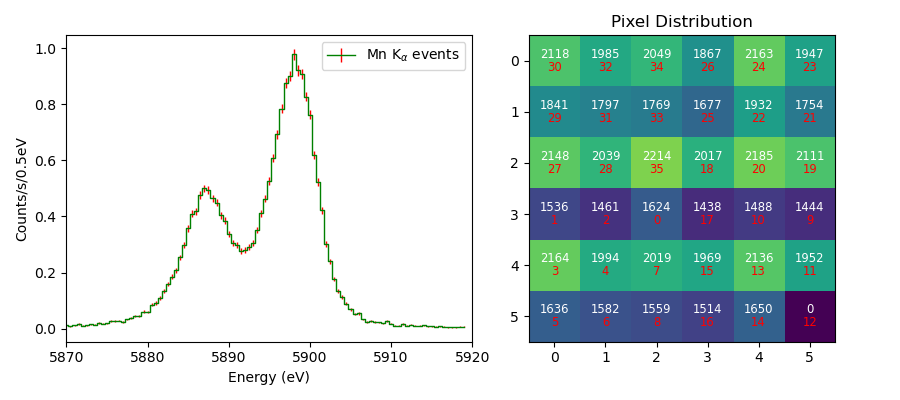}
\end{center}
\vspace{-0mm}
\caption{Illumination of resolve  {array} using the $^{55}$Fe filter position.   {The spectrum is that of the Mn K$_\alpha$ line at a resolution of about 4.5 eV.}  {The pixel number of the array is given in red and the total number of counts in white.}  Pixel 12 has been excluded from the map as this pixel is offset from the array proper  {and not illuminated by the $^{55}$Fe of the filter wheel.} \label{fig:fe55_line_dist}}
\end{figure}

When rotated to the $^{55}$Fe wheel position, the detector gain drifts can be monitored and calibrated.  5 radioactive sources of $^{55}$Fe are mounted on the four edges of the central cross and one in the center. The $^{55}$Fe isotope has a half-life of 2.737 years.    These sources illuminate the entire array with manganese (Mn) \xray{}s: Mn K$_\alpha$ and Mn K$_\beta$ \xray{}s.  The 5 samples on the FW are sufficient to calibrate the Resolve array in roughly 30 minutes\cite{porter_full}.  The appropriate cadence, the rate for rotating to the $^{55}$Fe position and obtaining fiducial measurements, has been developed\cite{porter_full} and used with great success.   With these fiducial measurements, Resolve is able to reach a resolution of 4.5 eV\cite{porter_full}.
Fig. \ref{fig:fe55_line_dist} shows an example of the Mn K$_{\alpha}$ line spectrum taken with Resolve in orbit and the array illumination pattern.  
 {The tartan check pattern seen in the right hand image is due to the mesh support structure of the closed gate valve}\cite{10.1117/1.JATIS.7.2.028005}.  
In the gate-value closed configuration , periodically rotating the $^{55}$Fe filter into the line of sight of Resolve is sufficient to calibrate the gain history of the array\cite{porter_full}.

\subsection{trending }
\label{sec:trend}
With the gate-value closed, the $^{55}$Fe sources in the filter wheel provide the fallback drift correction\cite{porter_full}.  It is imperative that the ability to rotate the $^{55}$ Fe sources into and out of the resolved beam is maintained.

All components of the FW have gone through extensive ground testing from the component level through to integration on the satellite.  The commissioning phase revealed nothing new.  The FW is working in flight as it did on ground.  

The filter wheel stepper motors are classified to operate for 90,000 60$^\circ$ or 10,000 360$^\circ$ rotations over Resolve's lifetime.  There are two open positions on the FW.  The OPEN1 position is only 60$^\circ$ from the $^{55}$Fe sources and currently is the default open position.  The health of the mechanism can be easily monitored by looking at housekeeping parameters: motor current and temperature of the motor housing.  Any degradation of the stepper motors is expected to show up as a significant variation from the trend as a dissipation of more heat for a given rotation.  

There are two FW motors, one prime and one redundant.  Each motor is a stepper motor with two phase windings.  Each winding has its own driver.  {T}he current for the prime motor is measured and downlinked as MOT1A and MOT1B in the FW housekeeping.  The temperature of the housing is measured near each redundant or prime motor.  

\begin{figure}[ht]
\begin{center}
\includegraphics[width=0.60\textheight]{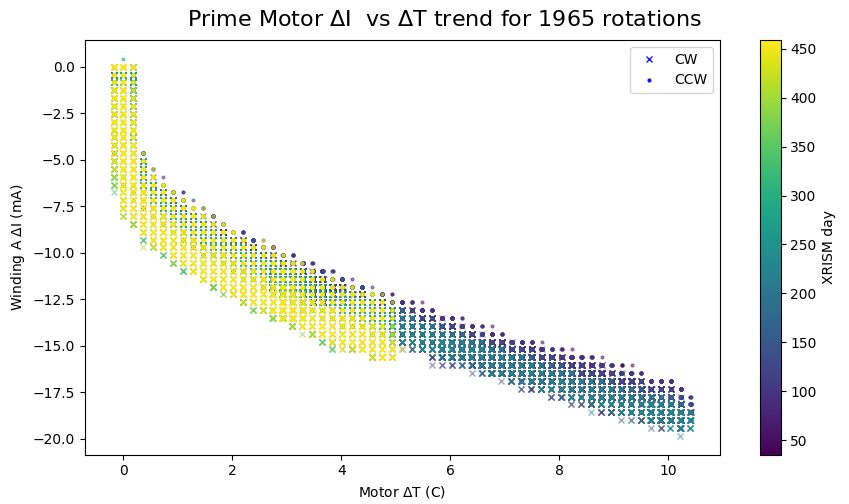}
\end{center}
\vspace{-0mm}
\caption{This plot shows the stepper motor A current versus temperature of the motor housing. The initial current and temperature are subtracted from each rotation to expose the very uniform trend. Each data point is a rotation step, either clockwise (CW) or counter clockwise (CCW).  The days since launch are color coded from dark blue just after launch to yellow for the most recent.  Rotations of 120$^\circ$ increase the temperature by roughly 10 C.  
In Feb 2024, the default open position was changed to OP1 which is 60$^\circ$ closer to the $^{55}$Fe source and results in a 5 C increase in temperature.
}
\label{fig:motortrend}
\end{figure}

Fig. \ref{fig:motortrend} shows the trend of current versus temperature of winding A of the prime stepper motor since the FW was first switched on in commissioning.   {Each point in the plot is a step of the motor, either clockwise (CW) or counter clockwise (CCW).}  {A rotation of 60$^\circ$ increases the temperature by about 5 C.}  The motor temperature changes naturally throughout an orbit, with orientation of XRISM, and  {when operating}.   {The current versus temperature trend is very consistent. In order to emphasize the consistency and show only the temperature change when operating the motors,} the difference from the initial current and temperature is calculated.  

Before 20 February 2024, the open position of the filter wheel (OP2) was used as the default open position.  The $^{55}$Fe position is a 120$^\circ$ rotation from OP2.  OP1 is only 60$^\circ$ rotation from $^{55}$Fe and therefore provides less wear to the motors.  The change was made in February 2024 to make OP1 the default open position and will reduce the overall wear of the motors. In Fig. \ref{fig:motortrend},  the points are color coded by day.  Before the change to OP1, rotation to and from 55Fe took 120$^\circ$ resulting in a temperature change of 10 C (blue/green points).  The most recent rotations appear in yellow. A temperature change of only 5 C is clearly visible in Fig. \ref{fig:motortrend}.  {Since the trend is very stable, } any signs of significant degradation should be easily identifiable.

\section{Modulated X-ray Source }
\label{sec:mxs}

There are two pairs (prime and redundant) of active calibration sources,  {MXSs}, on the calibration ring (see Fig. \ref{fig:test}B).  {To operate a MXS, a high voltage unit is used to supply voltages of a few to tens of kV (11.3 kV for Resolve), a light emitting diode (LED) and the MXS device itself which is a short vacuum tube with a photocathode at one end and layers of target metals (chromium and copper) on a beryllium window at the other.}
 {More specifically for Resolve, each MXS has a 25 $\mu m$ thick beryllium (Be) vacuum window. The photocathode is activated by blue/UV photons emitted by a LED. } 
 { The electrons produced by the photocathode after illumination by the LED are accelerated in the vacuum tube by 11.3 kV.} 
These electrons interact with a metal target deposited on the Be window to produce characteristic energy \xray{}s depending on the deposited metal layers.

The anode targets are chromium (Cr) and copper (Cu).  The direct MXSs target layers are 25 nm Cr on the Be window, 250 nm Cu on the Be/Cr, and another 50 nm Cr on the Cu layer.  This was a minor change from Hitomi-SXS MXS\cite{10.1117/1.JATIS.4.1.011204} to increase the efficiency of production of fluorescent Cu lines. The Hitomi/SXS MXS had 25 nm Cr on the Be window and 150 nm Cu on the Cr layer.  The direct MXS produces \xray{}s with characteristic energies for Cr (5.41 and 5.95 keV) and Cu (8.05 and 8.90 keV).  Additional fluorescence lines are visible:  {nickel} (Ni) at 7.6 keV and  {iron} (Fe) at 6.3 keV. These are caused by the Be window housing, which is also illuminated by the accelerated electrons.

Each pair of MXS is comprised of one direct MXS source which is oriented toward the Resolve array and one indirect MXS which does not directly illuminate the array but rather an aluminum/magnesium (Al/Mg) target just outside the Be window.  The Al/Mg target is oriented so that characteristic \xray{}s of Al (1.49 keV) and Mg (1.25 keV) will illuminate the Resolve array (see Fig. 12 of [\citenum{10.1117/1.JATIS.4.1.011204}]).
The  {prime} direct (MXS1) and indirect (MXS2) sources are mounted on the calibration ring, a quarter rotation around the ring from each other.     
The MXSs are fully redundant: both direct and indirect: The redundant sources are mounted in the calibration ring opposite each prime counterpart: redundant direct(MXS3) and redundant indirect (MXS4).

 {The MXS is designed to provide pulses of} \xray{}s  {of at known energies at specified times to illuminate the entire array.}  When the HV is at a high voltage (11.3 kV for Resolve), sending pulses of current to the LED results is a pulse of \xray{}s. The pulse provided to the LED of a MXS is highly tunable.  Three parameters determine the pulse: the pulse height, the pulse length, and the pulse spacing.  The pulse height is controlled by the current of the LED.  The pulse length  {is controlled by how long the LED is active and tunable} in steps of 0.125 ms up to 15.625 ms.  The pulse spacing  {determines the timing between pulses and} is tied to the onboard clock.  
Each SpaceWire tick is 1/64 sec (15.625 ms).  The maximum spacing is 4 sec.

Fixing the MXS pulses to the onboard clock has the additional benefit of aiding in the determination relative (see [\citenum{10.1117/12.2629753}]) and absolute timing (see [\citenum{sawada-timing}]) calibration of Resolve.

\subsection{gate valve closed}
 
\begin{figure}[ht]
  \begin{center}
  \includegraphics[width=1.0\linewidth]{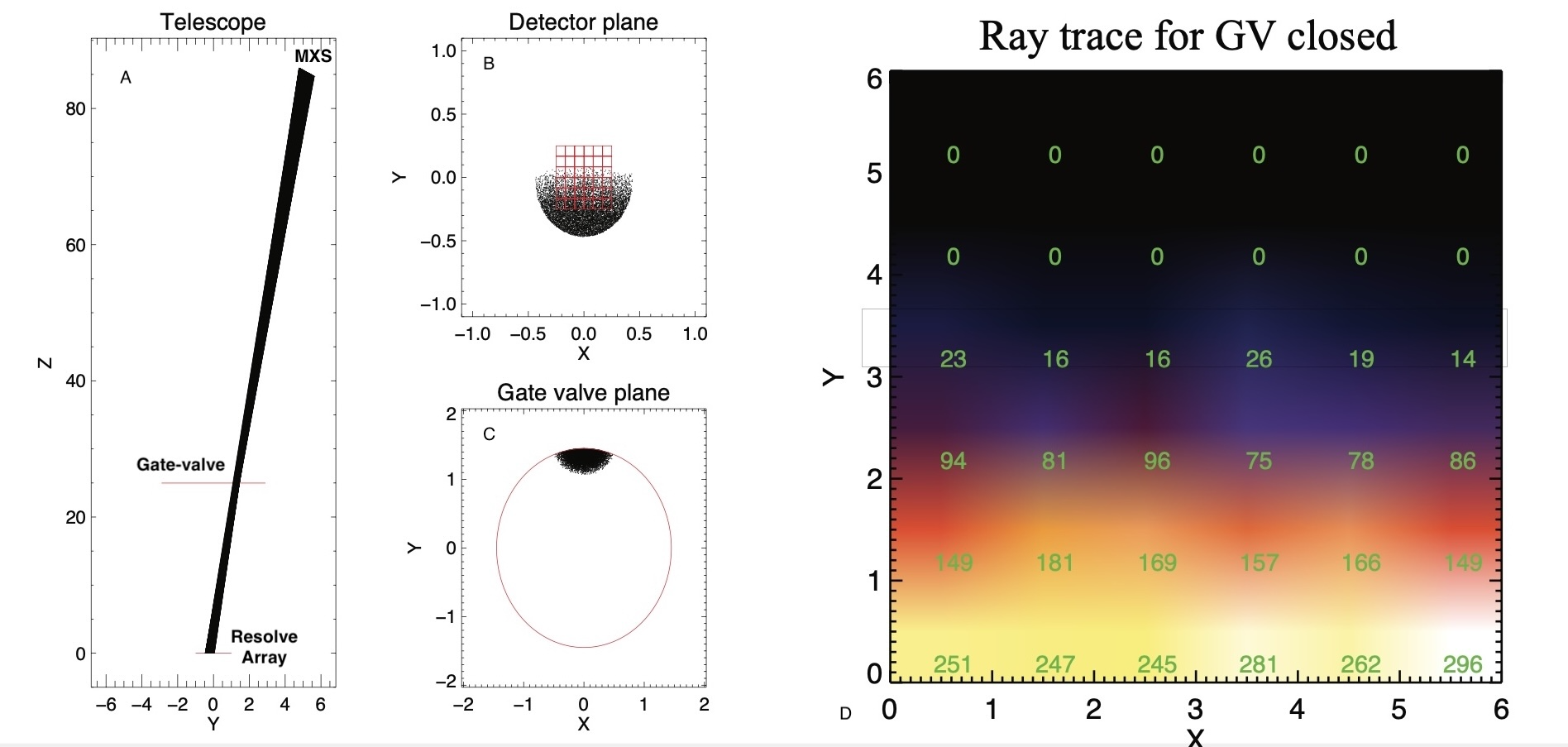}
  \caption{     
  Ray trace of gate valve blockage.  Panel A shows the geometry of a MXS and the Resolve array. 
  Each MXS is mounted roughly 90 cm above the array on the calibration ring (not shown).  Panels B and C show the how the \xray{}s from the MXS will fall in the detector plane and pass the gave-valve (respectively). The gate-valve sits on top of the dewar above the array.  Rays from the MXS that reach the array are shown as black dots.   Panel D is a prediction of the final illumination pattern in gate-valve closed configuration. Note: the orientation of illumination pattern is different with respect to Fig. \ref{fig:illumination}
  due to different convention of labeling of the X Y axis.}\label{fig:gvraytrace}
  \end{center}
\end{figure}

Fig. \ref{fig:gvraytrace}  {shows a ray trace of the illumination of the resolve array for a direct MXS when the gate-valve is closed. }
 {Panel A of Fig.} \ref{fig:gvraytrace}  {shows the geometry of a MXS with respect to Resolve's detector array and the gate-valve. Panels B and C show how the gate-valve partially blocks the} \xray{}s. \xray{}s  {which reach the array are shown as black dots. }  {Panel D is a prediction of the illumination pattern on an array of 36 pixels.}  {  Each MXS is partially blocked by the aperture holding the gate valve.}  Because the indirect source, MXS2, is mounted 90$^\circ$ around the calibration ring, the illumination pattern will also rotate by 90$^\circ$.  MXS3, the direct redundant MXS, is mounted on the opposite side of the calibration ring, 180$^\circ$ from MXS1.  In gate-valve closed configuration, all the Resolve pixels can be illuminated only by switching between prime (MXS1) and redundant (MXS3) not simultaneously.  {Note: }Fig. \ref{fig:gvraytrace}  {uses a different convention for labeling the X and Y axis and appears rotated relative to Fig.} \ref{fig:illumination}.
In a gate-valve open configuration, each MXS will fully illuminate the array.

\subsection{spectrum and array illumination}

\begin{figure}[ht]
  \begin{center}
  \includegraphics[width=1.0\linewidth]{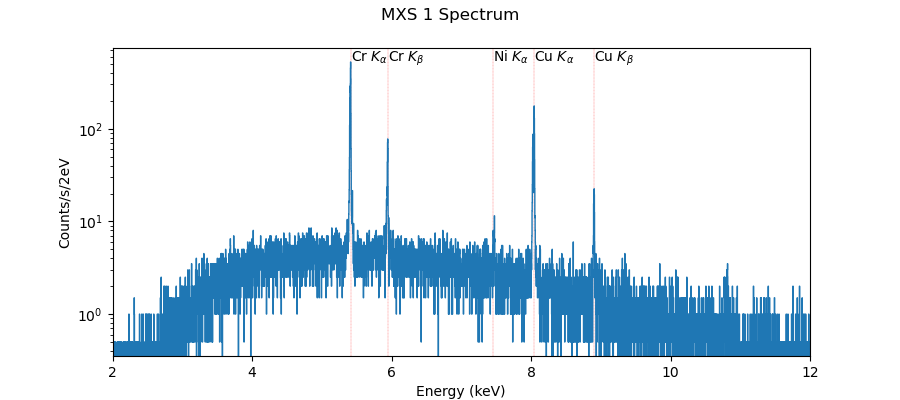}
  \caption{MXS1 (prime) in commissioning.  {This figure shows a standard spectrum of MXS1.  The pulse parameters are described in the text. The characteristic lines of Cr K$_\alpha$, Cr K$_\beta$, Cu K$_\alpha$ and Cu K$_\beta$ as well as Ni K$_\alpha$ are labeled.  The continuum is due to bremsstrahlung within the MXS device.} \label{fig:mxsspectrum}}
  \end{center}
\end{figure}

\begin{figure}

%
%
    \begin{center}
 
    \includegraphics[width=1.0\linewidth]{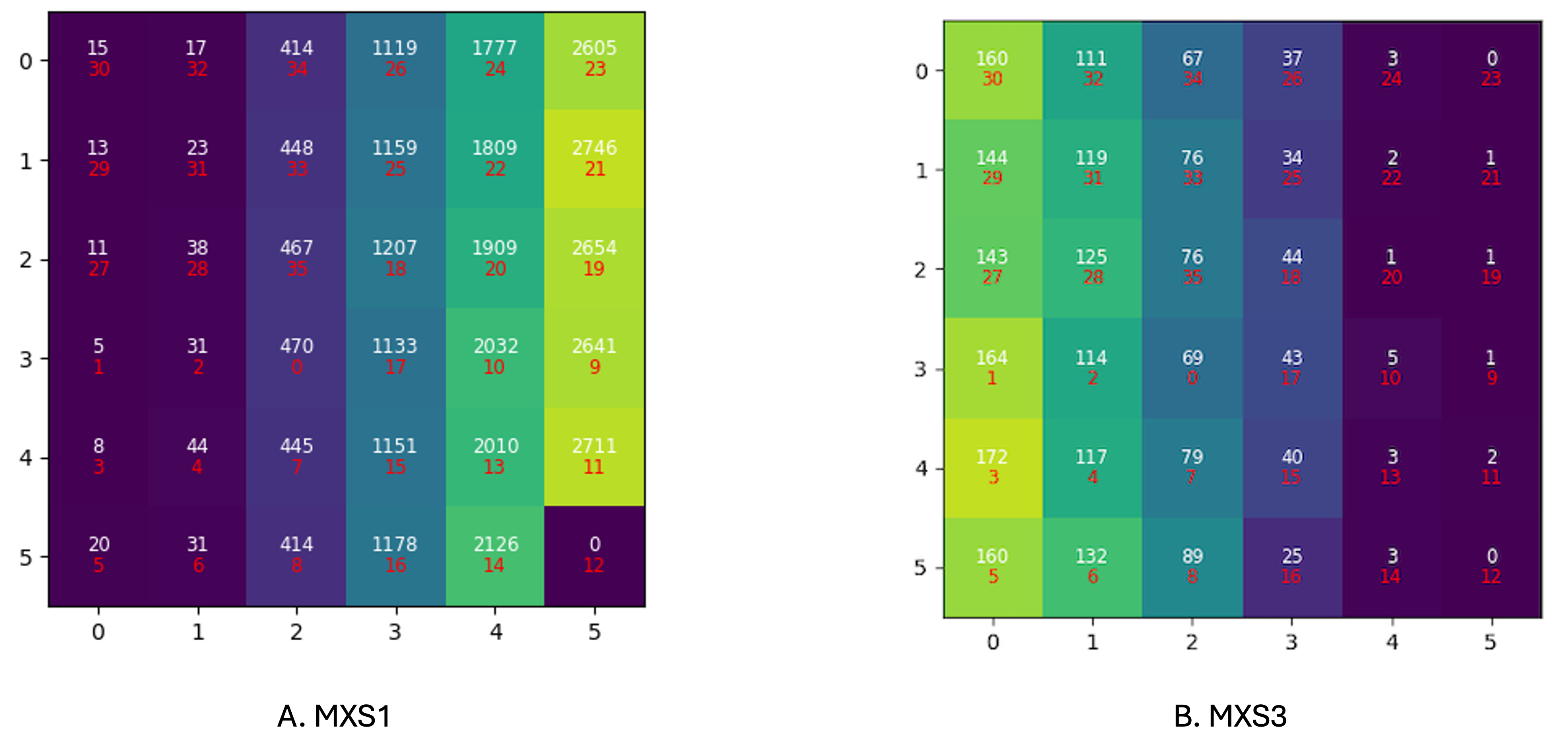}        
    \caption{Illumination pattern of Resolve pixels using prime direct MXS1 (Fig. \ref{fig:illumination}A) and redundant direct MXS3 (Fig. \ref{fig:illumination}B.  Resolve pixel numbers are indicated in red.  Total counts in each pixel is shown in white.  Events in the cal-pixel pixel 12 have not been included in the image map. 
           \label{fig:illumination}}
    \end{center}
\end{figure}

The commissioning of the prime direct MXS1 and indirect MXS2 started in late October 2023.  Fig. \ref{fig:mxsspectrum} shows a sample 1800-second spectrum while MXS1 / LED1 (prime,direct) was pulsed.   Resolve was pointed toward the blank sky.  {This spectrum is entirely consistent with prelaunch ground testing.} Fig. \ref{fig:illumination} shows the direct MXS1 and MXS3 illumination on the array.  In each pixel, the total counts (in white) of the pulse \xray{} are shown. The pixel ID is given in red.  This observation was taken just before the first attempt at opening the gate-vale, roughly half the array is illuminated due to shadowing of the gate valve. LED 1 was set at 1.02 mA for a pulse length of 0.625 ms at a pulse spacing of 93.75 ms.  Fig. \ref{fig:mxspulse} in Section \ref{sec:pulsetail} shows the pulse after folding the light curve.

In the spectrum of Fig. \ref{fig:mxsspectrum} the characteristic lines of Cr K$_\alpha$, Cr K$_\beta$, Cu K$_\alpha$ and Cu K$_\beta$ \xray{} as well as Ni K$_\alpha$ \xray{} are clearly visible.  The spectrum also shows the bremsstrahlung component.  The low-energy bremsstrahlung \xray{}s are stopped by the Be vacuum window, but energies above 2 keV create the 'continuum' of the MXS spectrum.

Roughly halve of the Resolve array is illuminated by MXS \xray{}s when the gate valve is closed, as shown in Fig. \ref{fig:illumination}A. 
Since the gate valve remains closed, the redundant direct source, MXS3, was commissioned in January of 2024 to illuminate the other side of the array.  Fig. \ref{fig:illumination}B shows the array map for the redundant direct source.   The illumination pattern is opposite to that of the prime MXS.  

The high voltage (HV) power supplies are also redundant, but the prime and redundant HV supplies are not designed to operate at the same time.  
The redundant source is on the opposite side of the calibration ring, and thus, using the prime and redundant source enables the full illumination of the array, but this is highly inefficient given the fact that the HV cannot be run simultaneously.

MXS2 and MXS4 are the indirect MXSs which are oriented to illuminate an aluminum plate (and magnesium) to produce low energy characteristic \xray{}s.  In general, due to blockage of the gate-valve to low energy \xray{}s, in closed gate valve configuration, the indirect source is not used.  The Be window of the gate valve blocks the lower energy lines of Al and Mg that the indirect source produces.
During early commissioning, the prime indirect source, MXS2, was activated during a standard switch-on procedure.  Due to elastic scatting off the Al target, the high-energy Cr and Cu \xray{}s do reach the array.  This also confirms the proper working of the indirect sources.  The illumination pattern on the array (not shown) is also rotated 90 degrees from the direct MXS1, consistent with the mounting in the calibration ring.

\subsection{pulse tail}
\label{sec:pulsetail}
\begin{figure}[tph]
\begin{center}
\includegraphics[width=0.50\textheight]{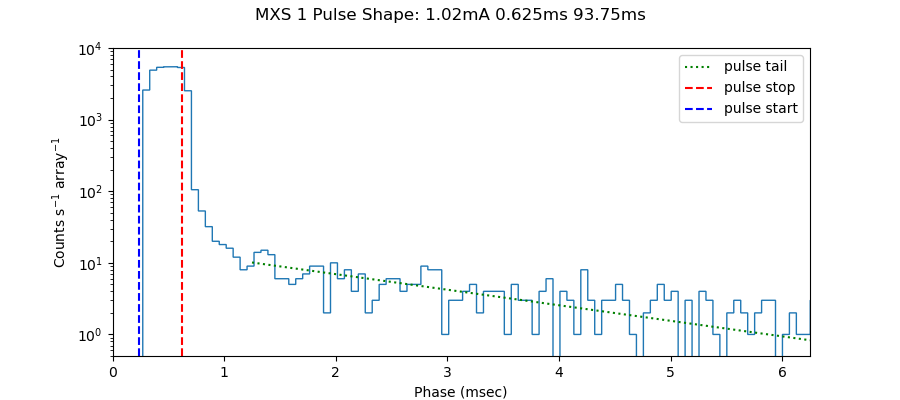}
\end{center}
\vspace{-0mm}
\caption{Example  {MXS1} pulse. This is a folded light curve folded every 93.75 ms.  Only the first 6 ms of the pulse are shown.  The pulse length is set to be 0.625 ms and the pulse height is determined by the LED current, 1.02 mA in this case. Dashed red and blue lines indicate the start time and stop time of the pulse.  The green dotted line shows enhance \xray{} emission after the pulse has stopped, i.e., the pulse tail.   \label{fig:mxspulse}}
\end{figure}

Fig. \ref{fig:mxspulse} shows a part of the folded light curve of the prime direct MXS1.  The pulse start and stop times are indicated as vertical dashed lines.  The pulse spacing (and hence folding) for these data was set to 93.75 ms.  The figure shows only the first 6 ms of the pulse to accentuate the slow decay of MXS1 produced \xray{}s after the pulse itself has finished: the pulse tail (dotted green line).   {The start and stop times of the pulse are indicated in the figure as blue and red vertical dashed lines}.  As observed in ground testing, the start time (blue dashed line in Fig. \ref{fig:mxspulse}) depends on the LED current.  The effect is well modeled and parameterized: higher currents starting closer to 0 sec.  The optimal pulse parameters to achieve the desired drift calibration in case the gate valve is open have been explored in [\citenum{sawada_full}]. 

 {Recent} investigations suggest that the tail is caused by photoluminescence of the MXS tube structure.  The \xray{}s produced in a MXS emit in a roughly isotropic distribution.  In particular, \xray{}s that hit the walls inside the MXS tube will be absorbed into the ceramic casing.  
These \xray{}s can cause luminescence, which in turn will activate the photocathode to release electrons\cite{akhil2024}.  This interaction produces an exponentially decaying afterglow of the X-ray pulse.  The time constants of the after glow range from a few to 10 ms\cite{ilija2022}.

\subsection{energy scale calibration}

The direct MXSs will be used to calibrate the energy scale.  The MXSs produces Cr K$_\alpha$ Cr K$_\beta$, Cu K$_\alpha$ Cu K$_\beta$ lines that span the energy range of 5.4 to 8 keV.  The results of those calibrations are discussed in [\citenum{megan_full}].  
 {In the current gate-valve closed configuration,} to use the direct MXSs to calibrate the entire array, dedicated observations are necessary, which use the prime and redundant direct MXSs in sequence. 
Unfortunately, as long as the gate valve remains closed, the lower-energy X-rays provided by the indirect source are not visible.   See [\citenum{megan_full}] for a discussion on extending the gain scale calibration to lower energies.

\subsection{light leak }

\label{sec:lightleak}

During ground testing of the FW flight spare on the spacecraft simulator at JAXA it was noted that the Resolve detector was receiving \xray{}s when turning on the HV to 11.3 kV but when the LEDs were still off.  Further investigation revealed that MXS1  {(prime / direct)} source was reacting to the ambient lighting of the test facility. The light from the environment illuminated the MXS1 cathode, releasing electrons inside the tube even when the LED was off.

Further testing identified entry points of the light leak and mitigation steps were carried out.  
These measures significantly reduced the scattered light that found its way to the photocathode, but validation in flight was needed.
  
\begin{figure}[tph]
\begin{center}
\includegraphics[width=0.70\textheight]{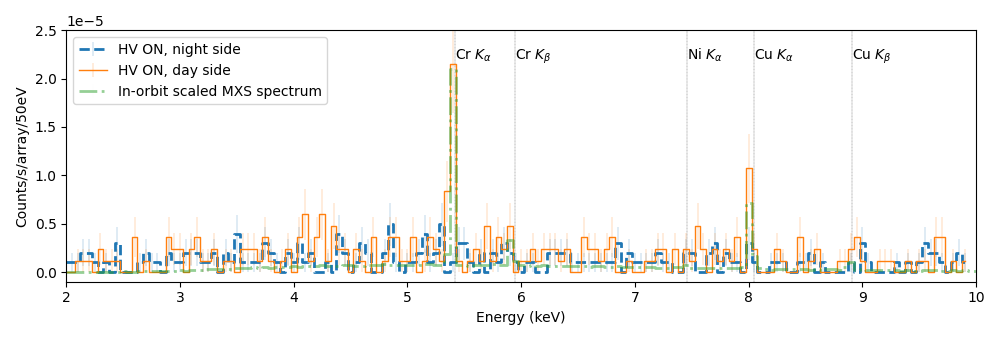}
\end{center}
\vspace{-0mm}
\caption{ {Light leak test results.  Blue-dash histogram (and error bar) is for all times when the high voltage (HV) was at 11.3 kV without pulsing the LED and XRISM was on the night side of the Earth (20209 seconds) , orange solid histogram are for times when the HV was on again without LED pulses and XRISM was on the sun side and exposed to reflection from the Earth (16777 seconds).   For comparison, the green dash-dot data are a standard spectrum when MXS1 was activated (HV on and LED pulsed) during early commissioning but scaled to match the Cr K$_\alpha$ line in orange.  Also indicated in this figure are the lines which will be present when operating an MXS.}   \label{fig:lightleak}}
\end{figure}

In its low Earth orbit, XRISM frequently passes between the Sun and Earth.  At these times, the Earth reflects sunlight onto XRISM.
Shortly after the first switch-on of the prime high voltage (HV1) in commissioning,  a few orbits with HV1 at 11.3 kV and LED1 not enabled were spent to test the light leak.   Fig. \ref{fig:lightleak} shows the results of the investigation.   {The blue histogram shows data when the HV is on, the LED is not pulsed and XRISM is on the night side of the Earth} in its orbit where no reflected light from the Earth can reach the calibration ring. The resulting spectrum shows just noise.   {However, the presence of Cr K$_\alpha$ and Cu K$_\alpha$ } \xray{}s  { (orange histogram) in the day-side only data indicate that reflected sunlight is reaching the photocathode activating the MXS and producing electrons.}   There is some indication that only specific times and spacecraft orientations with respect to the reflected sunlight activate the MXS, but this requires further investigation with more data.

Once identified during ground testing as a potential noise source, an allocation was made to be less than half of the non-\xray{} background (NXB) requirement. This corresponds to 0.5 c/s/keV.
Taking into account only day-side times, the measured flux of light leak \xray{}s is 2-3 times the allotment in specific energy bins around the Cr K$_\alpha$ and Cu K$\alpha$ lines.  In those energy bins, there would be an excess noise contribution due to the light leak.  

If the gate valve were to open, the Be vacuum window on the MXS effectively blocks energies below 2 keV, and hence the NXB background due to the light leak at these energies would be negligible.  The light leak for the redundant MXS3 has not been tested, but is likely to be different from what is reported here, given the different mounting of the redundant MXS3 on the calibration ring.

\subsection{MXSs in operations}

The original intended use of  {MXS1} was for the device to continually pulse during observations at a rate that is sufficient for drift correction.    The determination of the appropriate pulse parameters is key to efficiently using the MXS for gain tracking.  
The presence of unwanted MXS generated \xray{}s (pulse tail: Section \ref{sec:pulsetail} and light leak: Section \ref{sec:lightleak}) significantly complicates the finding of appropriate pulse properties for continuous gain tracking \cite{sawada_full}.  

When the gate valve eventually opens, [\citenum{sawada_full}] have proposed an alternative method of gain track using  {MXS1} periodically: intermittent mode.  This mode is similar to the current gain tracking with the position of the $^{55}$Fe filter.    In this mode, the prime direct MXS1 is used for short periods of time over an ADR cycle to create appropriate fiducials for gain drift calibration.  The time interval is excluded from the observation.  This intermittent mode avoids MXS generated \xray{}s (either from the pulse tail or the light leak) adding to the NXB.

\section{Conclusion}

The FW performs  {in} orbit as it did in ground testing. The FW was significantly tested and characterized prior to flight. Those tests resulted in a smooth transition to  {operations} in orbit. The filter wheel meets the requirements and allows good gain calibrations using the $^{55}$Fe sources on the filter wheel.

When the gate valve is opened, the MXS1 can be used for gain tracking, but only in an intermittent mode, much like how the $^{55}$Fe FW position is used now: switch-on for short periods of time to track the gain changes and then off to avoid the light-leak X-rays and afterglow tails. In addition, it can be used to verify the energy scale below 2 keV using  {MXS2} indirect source.

\section{Acknowledgments}

This article is based on a paper presented in the SPIE Astronomical Telescopes and Instrumentation conference, 16–21 June 2024, in Yokohama, Japan. \cite{10.1117/12.3017279}

Part of this work was performed under the auspices of the U.S. Department of Energy by Lawrence Livermore National Laboratory under Contract DE-AC52-07NA27344.

This research has made use of data, software and/or web tools obtained from the High Energy Astrophysics Science Archive Research Center (HEASARC), a service of the Astrophysics Science Division at NASA/GSFC and of the Smithsonian Astrophysical Observatory's High Energy Astrophysics Division.

\section{Code, data, and materials availability}
Data from the commissioning phase used in this article are proprietary
and are not publicly available. However, the figures presented in this article, as well as a limited subset
of the underlying data, are available upon request at r.f.shipman@sron.nl.

\section{Disclosures}
The authors declare that they have no financial interests, commercial affiliations, or other potential conflicts of interest that could have influenced the objectivity of this research or the writing of this paper.

\bibliography{fwinflight.bib} 
\bibliographystyle{spiebib} 

\end{document}